\documentclass{article}

\usepackage[english]{babel}

\usepackage[letterpaper,top=2cm,bottom=2cm,left=3cm,right=3cm,marginparwidth=1.75cm]{geometry}

\usepackage{amsmath,amssymb}
\usepackage{graphicx}
\usepackage[colorlinks=true, allcolors=blue]{hyperref}
\usepackage{cite}
\usepackage{enumitem}
\usepackage{helvet}
\usepackage{multirow}
\usepackage{authblk}
\usepackage[font=small,labelfont=bf]{caption}
\usepackage{subcaption}
\usepackage{xcolor}
\title{Axial Seamount Eruption Forecasting Experiment}
\author[1*]{Qinghua Lei}
\author[2]{Didier Sornette}
\author[3]{William W. Chadwick Jr.}
\author[4]{Scott L. Nooner}
\author[5]{Maochuan Zhang}
\author[5]{William S. D. Wilcock}

\affil[1]{Department of Earth Sciences, Uppsala University, Uppsala, Sweden}
\affil[2]{Institute of Risk Analysis, Prediction and Management, Academy for Advanced Interdisciplinary Studies, Southern University of Science and Technology, Shenzhen, China}
\affil[3]{Hatfield Marine Science Center, Oregon State University, Newport, USA}
\affil[4]{Department of Earth and Ocean Sciences, University of North Carolina Wilmington, Wilmington, USA}
\affil[5]{School of Oceanography, University of Washington, Seattle, USA}

\begin{document}
\maketitle
\vspace{-3em}

\begingroup
\renewcommand\thefootnote{}\footnote{
* Correspondence: \texttt{qinghua.lei@geo.uu.se}}
\renewcommand\thefootnote{}\footnote{
\textbf{Author contributions.}
Conceptualisation: Q.L. and D.S.;
Experiment design: Q.L., D.S., and W.W.C.;
Code development: Q.L. and D.S.;
Data acquisition: W.W.C., S.N., W.S.D.W., and M.Z.;
Forecast analysis: Q.L. and D.S.;
Document writing: Q.L., D.S., W.W.C., and M.Z.;
Scientific discussion: all authors.}
\addtocounter{footnote}{0}
\endgroup

\begin{abstract}
We introduce the Axial Seamount Eruption Forecasting Experiment (EFE), a real-time initiative designed to test the predictability of volcanic eruptions through a transparent, physics-based framework. The experiment is inspired by the Financial Bubble Experiment, adapting its principles of digital authentication, timestamped archiving, and delayed disclosure to the field of volcanology. The EFE implements a reproducible protocol in which each forecast is securely timestamped and cryptographically hashed (SHA-256) before being made public. The corresponding forecast documents, containing detailed diagnostics and probabilistic analyses, will be released after the next eruption or, if the forecasts are proven incorrect, at a later date. This procedure ensures full transparency while preventing premature interpretation or controversy surrounding public predictions. Forecasts will be issued monthly, or more frequently if required, using real-time monitoring data from the Ocean Observatories Initiative’s Regional Cabled Array at Axial Seamount. By committing to publish all forecasts, successful or not, the EFE establishes a scientifically rigorous, falsifiable protocol to evaluate the limits of eruption forecasting. The ultimate goal is to transform eruption prediction into a cumulative and testable science founded on open verification, reproducibility, and physical understanding.
\end{abstract}

\section{Introduction}
This document describes the Axial Seamount Eruption Forecasting Experiment (EFE), launched within the Geohazards Crises Observatory (\url{www.geohazards-observatory.com}) to test and advance physics-based approaches for the prediction of volcanic eruptions. The design of the EFE is inspired by, and emulates, the Financial Bubble Experiment~\cite{Sornette2009_FBE,Sornette2010a_FBE,Sornette2010b_FBE,Woodard2010_FBE}, which pioneered a transparent, real-world testing framework for forecast validation in the context of financial crises. The motivation for the EFE arises from the limitations of existing methods in operational volcano forecasting. Despite remarkable advances in high-precision, real-time monitoring, there remains no universally accepted or reliable framework for diagnosing and forecasting eruptions as they unfold. A prevailing view within both academic and operational communities is that long-term eruption forecasts are generally not feasible. An important objective in volcanology is therefore to develop physics-based models that provide a sound scientific foundation for eruption forecasting~\cite{NASEM2017_VolcanicEruptions}.

The Eruption Forecasting Experiment (EFE) aims to rigorously test the following two hypotheses.

\begin{itemize}[itemsep=2pt, parsep=0pt, topsep=2pt, partopsep=2pt]
    \item \textbf{Hypothesis H1:} Imminent volcanic eruptions can be forecasted in real time by diagnosing characteristic precursory patterns that reveal the system’s approach to catastrophic failure.
    \item \textbf{Hypothesis H2:} The timing of volcanic eruptions can be probabilistically forecasted with a reliability exceeding that of chance, within well-defined confidence bounds.
\end{itemize}

The Eruption Forecasting Experiment (EFE) is based on a physics‐based failure forecast model that has previously been applied only retrospectively to volcano monitoring datasets~\cite{Lei2025_CEE,Lei2025_EPSL}. This experiment represents the first application and testing of the model using real‐time data prior to an eruption, providing an opportunity to objectively evaluate its forecasting performance. Because retrospective analyses are susceptible to various forms of bias—including data snooping, hindsight reinterpretation, and post hoc model adjustment—the EFE is designed as a real‐time forecasting experiment to minimise such biases. Only forecasts that are documented and timestamped in advance of the events can provide a truly unbiased test of predictive capability.

The EFE implements a transparent protocol for digital authentication and delayed disclosure of forecasts. Each forecast is securely timestamped and its content cryptographically hashed using the \texttt{SHA-256} algorithm on a GNU/Linux platform. The resulting hash digests are made public immediately, while the complete forecast documents are released only after the predicted event has occurred. This approach ensures scientific integrity by providing verifiable proof of forecast issuance and eliminating any possibility of retrospective modification or model adjustment, thereby establishing a rigorous basis for evaluating predictive performance.

This project applies and extends a suite of physics-based models to diagnose critical transitions in volcanic systems. These methods, developed and tested over the past decade in the broader context of rupture and failure forecasting across diverse geophysical systems \cite{Sornette1995,Saleur1996,Sornette1998,Lei2025_CEE,Lei2025_EPSL,Lei2025_GRL,Lei2025_IJRMMS_Singularity}, are here applied to the Axial Seamount volcano. The site provides an exceptional natural laboratory owing to its dense instrumental network \cite{Chadwick2012_NatGeo,Nooner2016_Science,Wilcock2016_Science,Chadwick2022_G3}, which includes geodetic measurements of seafloor uplift and continuous seismic monitoring that together capture the deformation and seismic activity. The combination of rich observational datasets and advanced diagnostic tools enables a rigorous, physics-based evaluation of volcanic eruption predictability. The EFE team is continuously developing and validating the forecasting approaches, which will be progressively implemented and enhanced in future releases.

By publicly documenting all forecasts, whether successful or not, the EFE aims to establish a reproducible framework for evaluating eruption predictability. Our goal is to build a transparent and cumulative approach to volcano forecasting, grounded in physical understanding, statistical testing, and open scientific scrutiny. While we fully intend to release all results, we will do so only after the next eruption has occurred. This delay prevents premature speculation about the forecasts or methods and protects the integrity of the experiment until its performance can be objectively assessed. In the meantime, the ongoing monitoring allows us to evaluate how the model performs in real time and, if necessary, to make carefully documented adjustments to improve its robustness before the eruption. We also believe that presenting a complete set of forecasts together, rather than releasing them one by one, offers a clearer picture of the overall performance of our approach. Finally, to be convincing, the experiment must report every case—both successes and failures—so that its predictive skill can be judged without bias.

\section{Experimental Design}
Our proposed protocol for the experiment is as follows:

\begin{itemize}[itemsep=2pt, parsep=0pt, topsep=2pt, partopsep=0pt]
    \item \textbf{Forecast release schedule.} We will issue forecasts on a regular schedule announced in advance on the website of the Geohazards Crisis Observatory (\url{www.geohazards-observatory.com}). The experiment was launched with its first release on 8 November 2025, followed by approximately monthly updates. The current document constitutes the twelfth release, based on data available through 28 October 2026. We aim to provide forecast updates once a month, or more frequently as conditions warrant.
    
    \item \textbf{Continuous monitoring and analysis.} We plan to continuously analyse real-time monitoring data from Axial Seamount, including geodetic measurements of seafloor uplift and continuous seismicity records, from the Ocean Observatories Initiative (OOI) Regional Cabled Array (RCA), funded by the US National Science Foundation \cite{Kelley2014_MarineGeology}. Together, these data capture the deformation and seismic activity associated with magmatic processes preceding eruptions.

    \item \textbf{Forecast preparation.} When a confident forecast is identified based on our diagnostic models, we summarise the result in a dedicated document (e.g. a \texttt{.pdf} file) containing the key diagnostics, model parameters, and probabilistic estimates of eruption timing.
    
    \item \textbf{Digital fingerprinting.} The forecast document will not be publicly released immediately. Instead, we will compute its digital fingerprint using a modern cryptographic hash function (\texttt{SHA-256}) under a GNU/Linux environment, in compliance with current recommendations by the National Institute of Standards and Technology. This produces a unique alphanumeric signature for the file, such that any modification of the content would yield a different hash value, thereby ensuring integrity verification (see \ref{appendix:crypto} for a detailed description of the procedure). For long-term authenticity and proof of timing, the hash can optionally be submitted to a trusted time-stamping authority (RFC 3161) or anchored to an append-only public ledger (e.g. OpenTimestamps).
    
    \item \textbf{Meta-document creation.} We will maintain a master “meta-document” (this document), which includes a brief description of the methods and datasets, the SHA-256 hash of each forecast file, and the corresponding forecast issuance date. Each version of the meta-document is uploaded to a publicly accessible archive (e.g. \texttt{arXiv.org}), thereby generating an independent timestamp issued by a trusted third party. The initial version (v1) establishes the official start date of the experiment.

    \item \textbf{Versioned releases.} As new forecasts are generated, we will update the meta-document with the corresponding SHA-256 hashes and issuance dates, and upload it again to the same repository. The server automatically creates version numbers (v2, v3, etc.) while preserving all previous versions to ensure transparency. Each new version contains the cumulative list of all previous SHA-256 hashes and publication dates.

    \item \textbf{Final disclosure.} After the next eruption has occurred, we will publish the complete set of forecast documents in an open-access repository (e.g. \texttt{arXiv.org}) and on the website of the Geohazards Crisis Observatory (\url{www.geohazards-observatory.com}). The corresponding hash digests will be cross-verified to demonstrate consistency with the original hashes recorded in the meta-document. A detailed description of the document authentication procedure is provided in \ref{appendix:crypto}. A summary of all forecasts and their outcomes will be included in the final release.
\end{itemize}
\noindent This rigorous protocol ensures two key principles of scientific integrity:
\begin{enumerate}[itemsep=2pt, parsep=0pt, topsep=2pt, partopsep=2pt]
    \item All forecasts, successful or not, will ultimately be revealed.
    \item No selective reporting or modification of forecasts is possible, since each version is permanently timestamped and publicly archived.
\end{enumerate}

Once the forecast documents are made public, we will evaluate their quality and assess how the results test the two hypotheses introduced above. This will involve developing statistical criteria to define what constitutes a successful diagnosis of an impending eruption and a successful probabilistic forecast of its timing. The long-term goal is to develop, test, and improve physics-based models to aid in effective eruption forecasting.

\section{Monitoring Datasets}
\subsection{Site description}
\begin{figure}[htbp]
    \centering
    \includegraphics[width=0.8\linewidth]{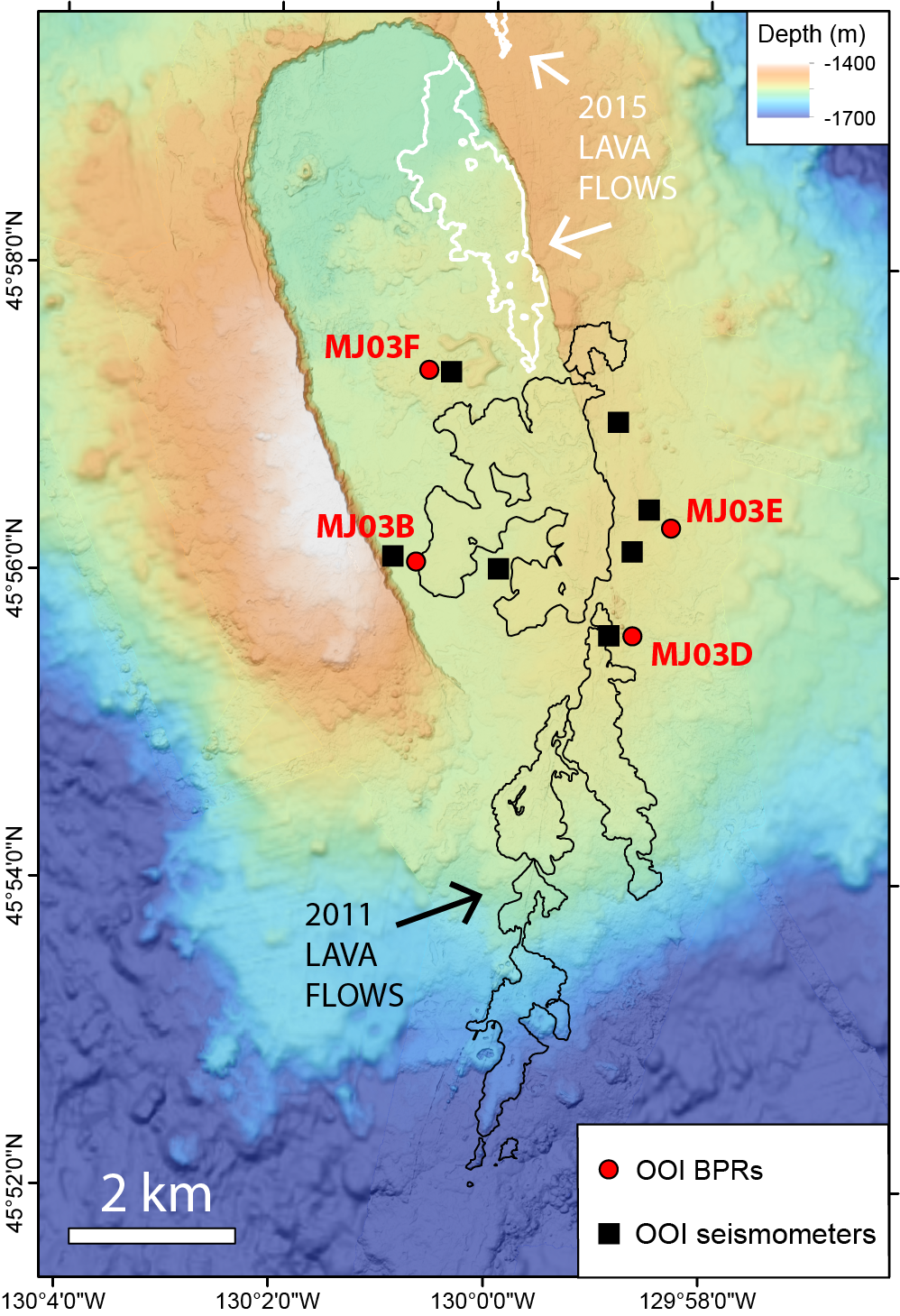}
    \caption{
        Bathymetric map of the summit caldera of Axial Seamount showing the network of cabled bottom pressure recorders (BPRs; red circles) and seismometers (black squares) installed as part of the Ocean Observatories Initiative (OOI) Regional Cabled Array (RCA).
    }
    \label{fig:axial_seamount}
\end{figure}

Axial Seamount is a submarine basaltic volcano located approximately 500~km offshore Oregon, USA~\cite{Chadwick2022_G3}. It has experienced three eruptive episodes over the past three decades, in 1998, 2011, and 2015. The volcano has been continuously monitored by a network of seafloor instruments~\cite{Nooner2016_Science}, and since 2014 by the OOI-RCA, which includes bottom pressure recorders, tiltmeters, and seismometers~\cite{Kelley2014_MarineGeology} (see Fig.~\ref{fig:axial_seamount}). This unique geophysical dataset provides an unprecedented opportunity to investigate pre-eruptive unrest processes and to test quantitative models for eruption forecasting. The 2015 eruption was previously anticipated about seven months in advance, within a one‐year window, based on pattern recognition in the geodetic time series~\cite{Nooner2016_Science}. This forecast was formulated empirically from the observed inflation pattern rather than derived from a physics‐based model. However, subsequent applications of this technique have not produced reliable forecasts due to variable inflation rates~\cite{Chadwick2024_AGU}, highlighting the need for physics‐based forecasting frameworks~\cite{Acocella2024_NatRevEarthEnv}. Within the current EFE project, we aim to perform ex-ante prospective forecasting of the next eruption of this volcano using our new physics-based failure forecast model with the continuous monitoring datasets of seafloor uplift and seismicity. All datasets in this version include observations updated to 28~October~2026.

\subsection{Seafloor uplift data}
\begin{figure}[htbp]
    \centering
    \includegraphics[width=1\linewidth]{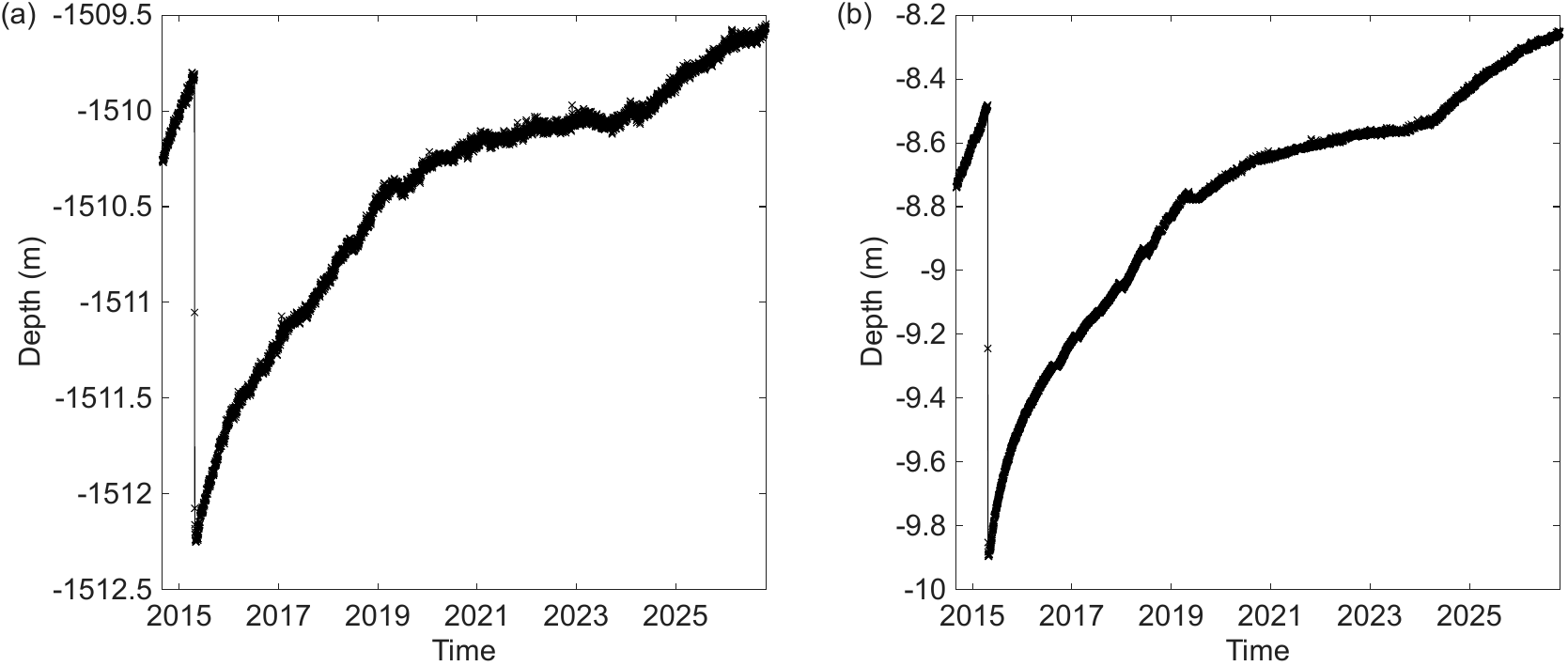}
    \caption{
        Seafloor uplift measurements at Axial Seamount were obtained from the OOI–RCA bottom pressure recorder (BPR) data repository \cite{Nooner2016_Science,Chadwick2022_G3} (updated as of 28~October~2026). (a) Time series recorded by the bottom pressure recorder (BPR) at the caldera centre (MJ03F in Fig. \ref{fig:axial_seamount}). (b) Differential BPR data derived by subtracting measurements at the reference site east of the caldera (MJ03E in Fig. \ref{fig:axial_seamount}) from those at the caldera centre (MJ03F). The major offset on the left side of the plots is the deflation that occurred during the 2015 eruption.
    }
    \label{fig:seafloor_uplift}
\end{figure}

Seafloor uplift measurements at Axial Seamount are obtained in real time from a network of four high-precision bottom pressure recorders (BPRs) deployed in and around the caldera~\cite{Chadwick2022_G3}. The cabled instruments record absolute pressure at a rate of 1~Hz. The raw pressure data are converted to depth and then de-tided using harmonic tide models. The drift of the OOI-RCA pressure sensors is constrained to less than 0.5~cm~yr$^{-1}$, based on comparisons with self-calibrating BPR instruments and remotely operated vehicle-based relative pressure measurements referenced to a stable site \cite{Sasagawa2016_ESS,Sasagawa2021_AGU,Chadwick2022_G3,Sullivan2025_AGU}. To suppress oceanographic noise (with amplitudes of about $\pm$5~cm), differential BPR records are also computed by subtracting measurements at a reference station located outside the caldera (MJ03E in Fig. \ref{fig:axial_seamount}) from the station at the center of the caldera (MJ03F in Fig. \ref{fig:axial_seamount}), reducing the residual noise to about $\pm$1~cm and better isolating the true geophysical signal of volcanic inflation and deflation \cite{Chadwick2022_G3}. It is important to note, however, that the differential BPR data is relative (MJ03F relative to MJ03E), and it likely includes some (small) component of slip across the seismically active eastern caldera fault. Fig.~\ref{fig:seafloor_uplift} shows the temporal evolution of seafloor displacement inferred from the single-station BPR record at the centre of the caldera and from differential BPR measurements.

\subsection{Seismicity data}
Seismic data are recorded by the OOI-RCA at Axial Seamount \cite{Wilcock2016_Science}. Earthquake detection is performed automatically using a short-term/long-term root-mean-square trigger algorithm applied to traces high-pass filtered at 4~Hz. Triggered signals from at least eight channels across four stations are grouped into a single event within a 1.5~s window to ensure robust multi-station detections. Automated classifiers exclude fin-whale calls and (co-eruption) seafloor explosions based on spectral ratios and waveform envelope correlations. P- and S-phase arrivals are refined using polarisation criteria and a kurtosis-based picker, and events are initially located with \textsc{HYPOINVERSE} (Klein, 2022) using a one-dimensional velocity model derived from caldera tomography. The formal absolute location uncertainties are typically $\sim$0.2~km horizontally and $\sim$0.3~km vertically.

\begin{figure}[htbp]
    \centering
    \includegraphics[width=1\linewidth]{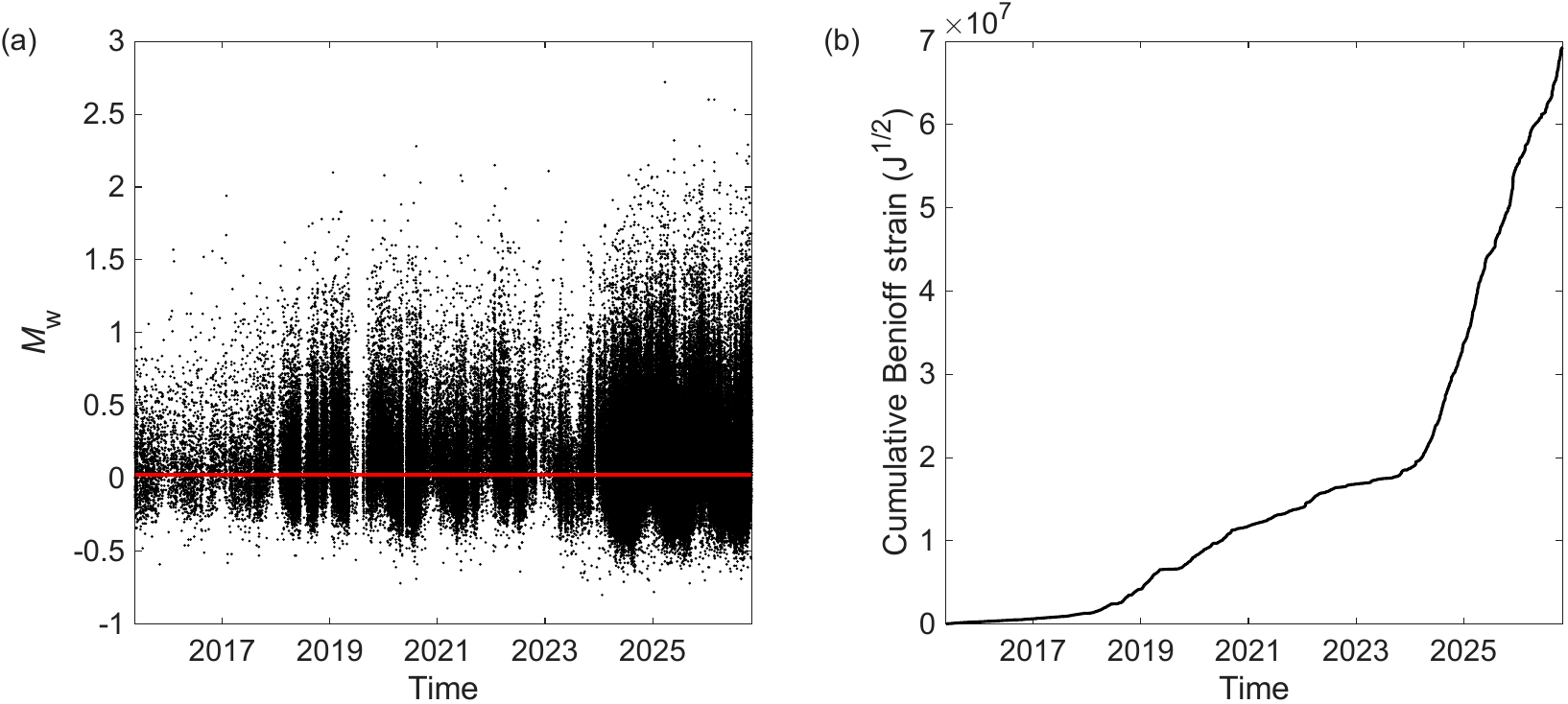}
    \caption{
        Seismic measurements at Axial Seamount from the OOI-RCA Earthquake Catalogue \cite{Wilcock2016_Science,Wilcock2017_IEDA} (updated as of 28~October~2026). (a) Magnitude-time sequence of earthquakes at Axial Seamount; the red line indicates the completeness magnitude of the catalogue. (b) Time series of cumulative Benioff strain derived from equation (\ref{eq:Omega_E}) with \(q = 1/2\).
    }
    \label{fig:seismic_data}
\end{figure}

For each event, P- and S-wave seismic moments $M_0$ were estimated from phase amplitudes following established techniques for marine microearthquakes, and the moment magnitude $M_\mathrm{w}$ was computed as \cite{Hanks1979_JGR}:
\begin{equation}
M_\mathrm{w} = \frac{2}{3}\log_{10}M_0 - 10.7,
\label{eq:Mw}
\end{equation}
\noindent where $M_0$ is in dyne$\cdot$cm. Moment magnitudes range from $-1.7$ to $3.4$, with an average of $0.2$. The catalogue magnitude is calculated as the median of the seismic moments measured across all contributing stations, ensuring consistent scaling among events. The magnitude of completeness $M_\mathrm{c}$ is estimated to be 0.05 based on the maximum curvature method applied to the Gutenberg-Richter frequency-magnitude distribution \cite{Wiemer2000_BSSA}. Fig. \ref{fig:seismic_data}a shows the magnitude-time sequence of earthquakes at Axial Seamount after the 2015 eruption.

We extract those events with $M_\mathrm{w} \geq M_\mathrm{c}$ and then compute the cumulative measure of seismic energy release as:
\begin{equation}
\Omega_E(t) = \sum_{i=1}^{N(t)} E_i^{\,q},
\label{eq:Omega_E}
\end{equation}
where the energy release \( E \) of the $i$th earthquake is estimated from \(\log_{10} E = 1.5\,M_\mathrm{w} + 4.8\) \cite{Kanamori1977_JGR} and \( N(t) \) is the cumulative number of earthquakes up to time \( t \). The power exponent \( q \) controls the relative weighting of small versus large events: \( q = 0 \) corresponds to the cumulative event count, \( q = 1/3 \) yields a measure proportional to the fault length, \( q = 1/2 \) gives a quantity scaling with the square root of seismic energy, \( q = 2/3 \) relates to the rupture area, and \( q = 1 \) leads to the cumulative energy release. Here, we adopt \( q = 1/2 \), corresponding to the so-called Benioff strain \cite{Bufe1993_JGR,Sornette1995}, with the resulting time series shown in Fig.~\ref{fig:seismic_data}b. Data gaps in the seismic catalogue were filled using piecewise cubic Hermite interpolation, which preserves the local trends without introducing artificial oscillations; tests with linear interpolation yielded similar results.

\section{Forecasting Methods}

\subsection{Theory}
Our theory for volcano forecasting builds on the log-periodic power-law singularity (LPPLS) formulation, which has proven effective in describing rupture dynamics in heterogeneous geophysical systems and has recently been validated, in retrospect, using a global dataset of 109 geohazard events \cite{Lei2025_CEE,Lei2025_EPSL,Lei2025_GRL,Lei2025_IJRMMS_Singularity}. The LPPLS model captures two essential ingredients of rupture dynamics: (i) the positive feedback mechanism, represented by a general power law acceleration that leads to a super-exponential trajectory culminating in a finite-time singularity, and (ii) the partial breakdown of continuous scale invariance into discrete scale invariance, which arises from the intermittent damage and rupture processes in heterogeneous materials. Importantly, the LPPLS model leverages the oscillatory patterns inherent in rupture dynamics, transforming what is traditionally perceived as noise into valuable predictive information. The first-order expansion of the general LPPLS formulation is given by \cite{Sornette1995,Lei2025_CEE}:
\begin{equation}
\Omega(t) \;=\; A \;+\; \bigl\{\, B + C \cos\!\bigl[\,\omega \ln(t_c - t) - \phi \bigr] \bigr\}\,(t_c - t)^{m},
\label{eq:lppls}
\end{equation}
\noindent where $\Omega(t)$ is the observable quantity (e.g. displacement, earthquake count, and seismic energy release), $t$ is time, $t_c$ is the critical time, $m$ is the critical exponent, $\omega$ is the angular log-periodic frequency, $\phi$ is a phase shift, and $A$ and $B$, and $C$ are constants. The LPPLS model embodies a log-periodic correction to the general power law acceleration of the system as it approaches eruption, capturing the varying rates of inflation observed during volcanic unrest. The critical exponent \( m \) must satisfy \( 0 < m < 1 \) for the deformation and seismicity to remain finite while their first derivatives diverge at the critical time \( t_c \), corresponding to a finite-time singularity. When the calibration of the LPPLS model yields \( m > 1 \), this should not be interpreted as the model describing a decelerating or stabilising regime, but rather as a diagnostic indication that no finite-time singularity is foreseeable within the current horizon of analysis. The fitted \( t_c \) does not correspond to an impending rupture but instead defines the horizon \( t_c - t_2 \) over which no meaningful forecast can be made. In this sense, the LPPLS framework remains valuable not for modelling stabilisation itself, but for revealing its own predictive limitation---indicating that the observed dynamics fall outside the accelerating regime that the model is designed to capture.

\subsection{Procedures}
The LPPLS model involves seven parameters, \( \boldsymbol{\theta} = \{ t_c,\, m,\, \omega,\, A,\, B,\, C,\, \phi \} \), with the critical time \( t_c \) being the primary parameter of interest. A stable and robust calibration scheme \cite{Filimonov2013_PhysicaA} is employed to fit the LPPLS model to the monitoring datasets of Axial Seamount in order to infer \( t_c \). For each monitoring dataset, the time series is extracted over the interval \([t_0,\, t_2]\), where \( t_0 \) is set to 19 May 2015, marking the end of the during-eruption period identified from deformation measurements~\cite{Nooner2016_Science} and earthquake focal mechanism analyses~\cite{Levy2018_Geology}, and \( t_2 \) corresponds to the current time when the forecast is made. The optimal starting time \( t_1 \) (\( t_0 \leq t_1 < t_2 \)) for LPPLS calibration is then determined using the Lagrange regularisation approach \cite{Demos2019_PhysicaA}, providing an estimate of the potential onset of the crisis (specific dates will be reported in the forecast results). The LPPLS model is subsequently calibrated using the data within \([t_1,\, t_2]\) to estimate \( t_c \) through a nonlinear fitting procedure \cite{Filimonov2013_PhysicaA}.

The monitoring data are aggregated on a weekly basis, with \( t_1 \) scanned in 7-day increments. For the uplift datasets (single-station and differential BPR data), daily averages are first computed, and then the median value within each 7-day interval is taken to represent the weekly observation, as the median is less sensitive to fluctuations and outliers. For the seismicity dataset (cumulative Benioff strain), the last recorded value within each 7-day interval is used to preserve the cumulative nature of the series. To ensure statistical robustness, at least 30 data points are required within the LPPLS calibration window, corresponding to a minimum duration of 210~days (\( t_2 - t_1 \geq 210~\mathrm{days} \)). The following parameter filters are also applied to enhance the robustness of the fitting: \( 0 < m < 2 \) and \( 2.5 < \omega < 25 \).

For probabilistic forecasting, three separate and complementary approaches are employed to derive the predictive distributions of \( t_c \), each based on different underlying principles:
\begin{itemize}[itemsep=2pt, parsep=0pt, topsep=2pt, partopsep=0pt]
    \item \textbf{Bootstrap.} Bootstrapping is performed by resampling the residuals from the ordinary least-squares LPPLS fit to generate multiple synthetic datasets, each of which is used to recalibrate the model. The resulting ensemble of \( t_c \) estimates defines the predictive distribution, which is obtained by applying an adaptive kernel density estimation to the ensemble \cite{Cranmer2001_KernelEstimation}.

    \item \textbf{Quantile regression.} Quantile regression is used to estimate the variability of the LPPLS fit across multiple quantile levels \cite{Zhang2016_PLOSOne}, capturing a range of possible trajectories consistent with the observation data. The ensemble of these quantile-based estimates forms another predictive distribution of \( t_c \), constructed through an adaptive kernel density estimation \cite{Cranmer2001_KernelEstimation}.

    \item \textbf{Modified profile likelihood.} The modified profile likelihood method constructs a likelihood function for \( t_c \) while integrating the effects of the other parameters. It provides a statistically rigorous way to estimate \( t_c \) and its confidence interval by explicitly accounting for the uncertainty of other parameters in the LPPLS model \cite{Filimonov2017_QuantFin}. The resulting likelihood function forms the basis for constructing the predictive distribution of \( t_c \).
\end{itemize}
Note that, for the bootstrapping approach, 20 ensemble realisations are generated to capture sampling uncertainty; for the quantile regression approach, the quantiles are scanned in 5\% increments, from 0.05 to 0.95, to construct the predictive distribution; for the modified profile likelihood approach, the confidence interval is estimated from the maximum of the modified likelihood function at the 5\% significance level. Each method thus yields a predictive distribution of \( t_c \), from which we report the median (50th percentile) together with the 25th and 75th percentiles. The three predictive distributions are compared to assess their consistency and subsequently synthesised to obtain an integrated forecast of \( t_c \).

Although these three approaches differ in implementation, they are all rooted in the same LPPLS theoretical framework and provide complementary measures of forecast uncertainty. Bootstrapping captures sampling uncertainty by generating many alternative realisations of the data through residual resampling. Quantile regression probes different parts of the residual distribution to reflect asymmetric or heterogeneous fluctuations \cite{Zhang2016_PLOSOne}. The modified profile likelihood focuses directly on how tightly the calibration can constrain the critical time~$t_c$, propagating parametric interactions and structural uncertainty into a likelihood-based confidence measure \cite{Filimonov2017_QuantFin}. Together, these methods illuminate distinct facets of uncertainty and yield a more complete characterisation of the predictive distribution of~$t_c$.

In addition to forecasting \( t_c \) using the latest data, we conduct a sensitivity analysis by systematically shifting \( t_2 \) backward in time to examine how the predictive distribution of \( t_c \) evolves with increasing data availability. In this analysis, the starting time \( t_1 \) of the LPPLS calibration window is fixed at the value determined from the latest dataset, which is based on the maximum amount of available data and is therefore expected to provide the best estimate of the onset of the crisis (so far).

We stress that our forecasting methodology should be viewed as an evolving framework rather than a finalised operational tool. Significant ongoing efforts aim to learn from practice through real-time testing, refine its formulation, improve robustness, and better quantify its limitations.

\section{Forecast Results}
Table~\ref{tab:efe_hashes} lists the cryptographic hash values (SHA-256) of the sealed forecast documents. Each hash uniquely authenticates the corresponding forecast version at the time it was issued.

\begin{table}[htbp]
\centering
\caption{SHA-256 cryptographic hash value of the sealed forecast document issued within the Axial Seamount Eruption Forecasting Experiment (EFE). The hash uniquely authenticates the corresponding timestamped forecast version.}
\label{tab:efe_hashes}
\small
\begin{tabular}{|l|l|l|}
\hline
Publication date & Document name & SHA-256 hash value \\ \hline
2025-11-08 & \texttt{efe\_001.pdf} & {\footnotesize\texttt{9eac807b0904db50530e9a6717004ea04ff3855c880aa04ef207e8e59034a328}} 
\\ \hline
2025-12-22 & \texttt{efe\_002.pdf} & {\footnotesize\texttt{f2ed5ade9a43f51edb874b474ebd8649622e2149e5fc5c63c88994f51d969d24}} \\ \hline
2026-01-30 & \texttt{efe\_003.pdf} & {\footnotesize\texttt{e01ad91530f0fece58fe6bcdc7aa2761227997e72209c7c8d384801d4ca8f8a0}} \\ \hline
2026-02-28 & \texttt{efe\_004.pdf} & {\footnotesize\texttt{ff46cc3c837bd44b615a3f6446a2336b90e51b37e7436a75f49e7109d396aedd}} \\ \hline
2026-03-31 & \texttt{efe\_005.pdf} & {\footnotesize\texttt{cc1481479dadfcf51720602211cb82cbdbc2c0ebb809b81a912286bc1542c7a2}} \\ \hline
2026-04-30 & \texttt{efe\_006.pdf} & {\footnotesize\texttt{91f890d02c77e5d4b1221f1423ae7a78dfea2ac5f99a27c95c645d4ffa831b1d}} \\
\hline
2026-05-31 & \texttt{efe\_007.pdf} & {\footnotesize\texttt{15e42096be15723b5c9bf9158b18a37b169cf44b6aa0fe24abcd408f479ca9a2}} \\
\hline
2026-07-01 & \texttt{efe\_008.pdf} & {\footnotesize\texttt{2427b7e29f4bb9b889112277d0a46319b88e55a24010ed982b4a7b3b33439c28}} \\
\hline
2026-08-04 & \texttt{efe\_009.pdf} & {\footnotesize\texttt{34c0106a8aaa9c5db0c759f20390af3da4fd46c8079e50be2cc8f5f195c44867}} \\
\hline
2026-09-08 & \texttt{efe\_010.pdf} & {\footnotesize\texttt{1cad4daa05814d0acb1179b9b2a651cdacec42c70a7e713b28b1249050dd4913}} \\
\hline
2026-11-01 & \texttt{efe\_011.pdf} & \footnotesize\texttt{5445243820018165014e1ea77e210f5533f931eae4c95e5053150f5c14983bc7} \\
\hline
2026-11-08 & \texttt{efe\_012.pdf} & \footnotesize\texttt{ff303ba8351005661be6d232f8a225ca8447fb4e755fbcd7883e7098845827e5} \\
\hline
\end{tabular}
\vspace{2pt}
\begin{minipage}{1\linewidth}
\footnotesize
\emph{Note:} SHA-256 belongs to the SHA-2 family of cryptographic hash functions (\url{https://en.wikipedia.org/wiki/SHA-2}).
\end{minipage}
\end{table}

The following output shows the version of the program used to generate the cryptographic hash values on our GNU/Linux system:

\noindent\rule{\linewidth}{0.4pt}
\begin{verbatim}
$ sha256sum --version
sha256sum (GNU coreutils) 9.4
Copyright (C) 2023 Free Software Foundation, Inc.
License GPLv3+: GNU GPL version 3 or later <https://gnu.org/licenses/gpl.html>
This is free software: you are free to change and redistribute it.
There is NO WARRANTY, to the extent permitted by law.

Written by Ulrich Drepper, Scott Miller, and David Madore.
\end{verbatim}
\rule{\linewidth}{0.4pt}

\appendix
\renewcommand{\thesection}{Appendix~\Alph{section}}
\section{Timestamping and Verification Procedure}
\label{appendix:crypto}

To ensure full transparency and prevent any possibility of retrospective modification, each forecast in the EFE is authenticated using a simple but robust cryptographic timestamping procedure. The aim is to demonstrate unambiguously that (i) a forecast document already existed at a specific date, and (ii) its contents have not changed since that time.

For each forecast, a confidential \texttt{.pdf} file is created and stored privately. A SHA-256 cryptographic hash of the file is then computed using a standard GNU/Linux utility such as:
\begin{verbatim}
sha256sum efe_001.pdf
\end{verbatim}
with the resulting 64-character hexadecimal digest serving as a unique digital fingerprint of the document. Any alteration to the file---even a single character, pixel, or whitespace---would produce a completely different hash, while reproducing the original document from the hash alone is computationally infeasible. The hash thus acts as a ``digital DNA'' fingerprint of the forecast document.

This hash value is recorded in the meta-document uploaded to the publicly accessible archive (e.g. arXiv.org). Because only the hash is revealed, the content of the forecast remains confidential, but its existence at the time of publication is cryptographically certified.

When all sealed forecast documents are released after the next eruption, anyone can independently verify their integrity by recomputing the SHA-256 hash of each document and comparing it with the previously published digest. Exact agreement provides definitive proof that the released file is bit-for-bit identical to the one that existed at the publicly recorded timestamp.

For convenience, we provide below the commands that readers may use to verify the
authenticity of the released forecast documents. Suppose the forecast file
\texttt{efe\_001.pdf} and its previously published hash file \texttt{efe\_001.sha256} have both
been downloaded. The integrity check is performed with:
\begin{verbatim}
sha256sum -c efe_001.sha256
\end{verbatim}
If the file is authentic and unmodified, the terminal will report:
\begin{verbatim}
efe_001.pdf: OK
\end{verbatim}
Alternatively, the file’s hash can be recomputed manually and compared directly with the value published in the meta-document:
\begin{verbatim}
sha256sum efe_001.pdf
\end{verbatim}
The resulting 64-character hexadecimal digest should match the published value exactly.

If optional GPG signatures are provided (a brief cryptographic stamp created using the
authors’ private key to prove authorship and authenticity), they may be verified as follows.
First, import the public key included in the forecast archive:
\begin{verbatim}
gpg --import efe_public_key.asc
\end{verbatim}
Then verify the signature associated with the hash file:
\begin{verbatim}
gpg --verify efe_001.sha256.asc efe_001.sha256
\end{verbatim}
A message such as:
\begin{verbatim}
gpg: Good signature from "Qinghua Lei <qinghua.lei@geo.uu.se>"
\end{verbatim}
confirms that the hash file is authentic and was signed by the forecast authors.

\section*{Acknowledgements}
This material is based upon work supported by the Ocean Observatories Initiative (OOI), a major facility fully funded by the U.S. National Science Foundation under Cooperative Agreement No.~2244833, and the Woods Hole Oceanographic Institution OOI Program Office. Q.L.\ is grateful for support from the European Research Council (ERC) under the European Union’s Horizon Europe programme (ERC Consolidator Grant, grant no.\ 101232311) for the project ``Unified framework for modelling progressive to catastrophic failure in fractured media (FORECAST)''. Q.L. and D.S. acknowledge partial support from the Norwegian Water Resources and Energy Directorate (NVE). D.S. acknowledges funding by the National Natural Science Foundation of China (Grant No. T2350710802 and U2039202), and the Center for Computational Science and Engineering at Southern University of Science and Technology. W.W.C. and S.N. acknowledge funding from US National Science Foundation awards (Grant No. OCE-2226488 and 2226445).

\bibliographystyle{unsrt}
\bibliography{references}

\end{document}